# A Tree Pattern Matching Algorithm for XML Queries with Structural Preferences*

## Maurice Tchoupé Tchendji[1], Lionel Tadonfouet[2], Thomas Tébougang Tchendji[1]

[1]Department of Mathematics and Computer Science, Faculty of Sciences, University of Dschang, Dschang, Cameroon
[2]INRIA, Paris, France
Email: ttchoupe@yahoo.fr, maurice.tchoupe@univ-dschang.org, lionel.tadonfouet@inria.fr, tttchendji@gmail.com

## Abstract

In the XML community, *exact queries* allow users to specify exactly what they want to check and/or retrieve in an XML document. When they are applied to a semi-structured document or to a document with an overly complex model, the lack or the ignorance of the explicit document model (DTD—Document Type Definition, Schema, etc.) increases the risk of obtaining an empty result set when the query is too specific, or, too large result set when it is too vague (e.g. it contains wildcards such as "*"). The reason is that in both cases, users write queries according to the document model they have in mind; this can be very far from the one that can actually be extracted from the document. Opposed to exact queries, *preference queries* are more flexible and can be relaxed to expand the search space during their evaluations. Indeed, during their evaluation, certain constraints (the preferences they contain) can be relaxed if necessary to avoid precisely empty results; moreover, the returned answers can be filtered to retain only the best ones. This paper presents an algorithm for evaluating such queries inspired by the *TreeMatch* algorithm proposed by Yao *et al.* for exact queries. In the proposed algorithm, the best answers are obtained by using an adaptation of the Skyline operator (defined in relational databases) in the context of documents (trees) to incrementally filter into the partial solutions set, those which satisfy the maximum of preferential constraints. The only restriction imposed on documents is No-Self-Containment.

---

*This paper is the extended version of the one titled *Evaluation des requêtes avec préférences structurelles sur les documents XML* [1] presented at CARI'14.

## 1. Introduction

Semi-structured documents are increasingly used in the IT community for publishing as well as for exchanging information among applications. Their ever increasing number coupled with the diversity of users and uses encourages the development of ever more efficient techniques for their usage in particular, through dedicated quering languages [2]. Indeed, as for traditional Databases (DB), the need to develop techniques for extracting information from a collection of XML documents (we are talking about XML DB) was very quickly felt, as well to meet the requirements of users/applications needing to explore the web or an XML DB, or for exploiting active documents[1] [3].

Initially, query techniques used in classical DBs had been adapted for use in XML DBs. Unfortunately, this approach is not always appropriate because it only allows to manage the XML subfamily of structured documents[2], leaving aside that made up of the large collection of the so-called semi-structured documents. In fact, as opposed to the data manipulated in classical DBs that are structured (we know the schema of the DB), those manipulated in the XML community are generally self-descriptive (semi-structured) *i.e.*, do not have a document model (DTD—Document Type Definition, Schema, etc.). This makes their exploitation and specifically their interrogation through queries very difficult, because of the inexistence or lack of knowledge (by the queries initiator) of their underlying structures: we want to query a document for which we do not know its exact structure (its model). Queries are formulated on the basis of what users believed about the content and/or the structure of the document to be interrogated. Usually, in such situations, in order to maximize chances of obtaining a convincing result, one formulates queries in the most general way to avoid as much as possible having an empty result. In doing so, by the extreme minimization of constraints, one can obtain a very important number of results.

The problems of the absence and/or overabundance of results for a query have also emerged within the community of relational DB (RDB). As a solution, the concept of *preference queries* was proposed and specific languages have been developed: SQLf [4], Preference SQL [5], Preference Queries [6], etc. This concept has also been adopted within the XML community and languages such as XPref [7], Preference XPath [8] etc. have been created.

Intuitively, a preference query specifies the user's desiderata and has two parts: a main condition (similar to a classical query; it is also called *exact query* or *strict query*) which aims to select for a RDB a set of n-tuples, or if it is an XML DB, all subtrees of the DB corresponding to the given (tree) pattern, and a preferential part used to specify optional (but preferred) requirements; these requirements can be quantitative or qualitative. Note immediately that an answer to a prefe-rence query may not satisfy all or some of the requirements specified in its preferential part, but must necessarily satisfy all those of the main condition.

---

[1]An active document is a document with dynamic parts (nodes) encapsulating queries on other documents. Their dynamic parts (nodes) are supplemented by necessity—at the moment of its use—by executing queries that they encapsulate.

[2]An XML document is said to be structured if it is valid with respect to a given document model; for example, a DTD, Schema, etc.

Initially, the concept of preference queries was motivated among others by the observation that DB users do not generally want to obtain all the answers of a query, but rather, the *best* or *most preferred ones*. Subsequently, it was applied in the opposite case (case of an empty answer) to return answers that are "close" to user's wishes; it's according to this understanding that they can be qualified as *flexible*.

Like a RDB, an XML document contains information (data); it also encapsulates a structure that must be taken into account during its querying. In order to transpose the preference concept to XML documents, Sara Cohen *et al.* in [9] distinguish two types of preferences: those concerning values (the user may prefer results that contain certain values, e.g. low price, best price, etc.) and those related to the document's structure (the user may prefer results with a certain structure, e.g. existence of a "discount" node or existence in the document's structure of an arc between "departure" and "arrival" nodes etc.).

The problem addressed in this paper concerns the evaluation of XML preference queries wherein preferences are related only to the document's structure. Our goal is to propose an algorithm for evaluating such queries by relying on the optimal algorithm proposed by Yao *et al.* in [10] for the evaluation of exact queries.

The proposed algorithm will use an adaptation of the *Skyline operator* (defined in RDB [11]) in the context of XML documents (trees), to incrementally filter the best answers from partial solutions of a query (tree). Finally, it will only return answers that satisfy a maximum of constraints called *SkyTrees* (see Sec. 3.3): they are *the best answers* since they are not dominated according to the Skyline operator.

In this paper, 1) we assume as in [10] that the used XML documents are non-recursive[3]: from this hypothesis, an important property is derived (see Sec. 2.2.1) that allow to considerably accelerate as in [10], the complexity of the proposed algorithm; 2) to clearly present our approach, we use a sub-language of the query language proposed by Sara *et al.* [9]; in fact, we focus only on preferences related to structure called in the following, *structural preferences*.

***Organization of the manuscript***: Section 2 presents some concepts related to semi-structured documents as well as the principle of the *TreeMatch* [10] algorithm. Section 3 is devoted to the presentation of *TreeMatchPreference*, an extension of *TreeMatch* to the case of preferences queries. Section 4 presents an experimental study of *TreeMatchPreference* as well as the prototype built for this purpose. Finally, Section 5 is dedicated to the conclusion.

## 2. Preliminaries

We present in this section some concepts related to semi-structured documents used in this manuscript. Concepts of *exact queries* and *preference queries* are defined more formally. Moreover, the principle of *TreeMatch* algorithm [10] allowing the evaluation of exact queries is also presented.

---

[3]In such documents, no element (Tag) is in the list of its descendants.

## 2.1. XML Documents, Queries and Evaluations

An XML document is a text file created according to the specifications of the XML standard[4]. From a purely organizational point of view, it consists of a set of logical structuring units called "elements" or "Tag"; it can be abstractly represented[5] by a tree $D = (N_d, E_d)$ where $N_d$ is a set of tagged nodes; each node of $N_d$ represents a Tag contained in the document and is tagged by this Tag. $E_d$ is a set of arcs; each arc connects two nodes of $N_d$ according to the existing relationship between tags involved in the (textual) document. For any node $x$ in $N_d$, function *labeld*($x$) returns its tag. For example, in **Figure 1(b)** we have a tree representation of an XML document in which a labelled node "*a*1" represents an occurrence of the "*A*" tag of the document.

As for XML documents, an *exact structural query Q* can also be represented by a tree $Q = (N_q, E_q)$ in which $N_q$ is a set of labelled nodes, $E_q$ a set of arcs each connecting two nodes of $N_q$; each arc is represented by the pair of nodes it connects. If *ar* is an arc, function *orig*(*ar*) (resp. *dest*(*ar*)) returns the source (resp. destination) node of *ar*. For example, if *ar* connects $n_s$ and $n_d$, we write $ar = (n_s, n_d)$, and therefore, *orig*(*ar*) = $n_s$ and *dest*(*ar*) = $n_d$. There are two types of arcs in $E_q$: those connecting a parent node to one of his children denoted *child* (*x, y*) (or *x/y*) and those connecting a node to one of his descendants denoted *desc* (*x, y*) (or *x//y*). For any node $x \in N_q$, function *labelq* (*x*) returns its label. **Figure 1(a)** presents a tree representation of an exact (structural) query containing only parent-child arcs.

Let, $Q = (N_q, E_q)$ be a query, $D = (N_d, E_d)$ an XML document, two nodes $n_d \in N_d$ and $n_q \in N_q$, two arcs $ad \in E_d$ and $aq \in E_q$:

- We'll say that, $n_d$ *is an occurrence of* $n_q$ in $D$ if *labeld* ($n_d$) = *labelq* ($n_q$). Similarly, we'll say that *ad is an occurrence of aq in D* if *orig* (*ad*) is an occurrence of *orig* (*aq*) and, *dest*(*ad*) is an occurrence of *dest*(*aq*).

- Let $\psi : N_q \mapsto N_d$ be an application. We'll say that $\psi(N_q)$ is a matching of $Q$ in $D$ if the following three conditions are satisfied for all $x, y \in N_q$: 1) $x = y \Leftrightarrow \psi(x) = \psi(y)$. 2) $label_q(x) = label_d(\psi(x))$. 3) if $x//y \in E_q \Rightarrow \psi(x)$ is an ancestor of $\psi(y)$; if $x/y \in E_q \Rightarrow \psi(x)$ is the parent of $\psi(y)$ *i.e.* $\psi(x)/\psi(y) \in E_d$.

- $Q$ *is satisfied in* $D$ if there is a match of $Q$ in $D$. An example of query matching is shown in **Figure 1** (blue lines).

As stated earlier, a preference query makes it possible to express requirements some of which are flexible *i.e.* may not be satisfied when matching. As for exact queries, a preference query can be represented by a tree $(N_q, E_q)$ in which: 1)

---

The set of nodes $N_q$ is partitioned into two subsets: $N_{req}$ containing the required or strict nodes—each of them must necessarily have an occurrence in any result of the query—and $N_{pref}$ containing the preferred nodes *i.e.* may not have an occurrence in the results. 2) The $E_q$ set of arcs is likewise partitioned into two subsets: $E_{req}$ containing required arcs and $E_{pref}$ containing preferential arcs.

A preference query $P = \left( N_{req}, N_{pref}, E_{req}, E_{pref} \right)$ is satisfied on a document $D$ if the exact query $Q_p = \left( N_{req}, E_{req} \right)$ is satisfied on $D$ and elements of $N_{pref}$ as well as those of $E_{pref}$ possibly have occurrences in $D$. **Figure 2(a)** is an example of a preference query's tree representation; nodes whose labels contain "?" are *preference nodes*. The blue lines in **Figure 2** show an example of a matching of a preference query on the document of **Figure 1(b)**. Note on this figure that all the required nodes (a, b, c, f) have occurrences; the preference node "d" has some too. On the other hand, the preference node "e" does not have an occurrence in the highlighted solution.

### 2.2. The *TreeMatch* Algorithm

Native evaluation approaches of XML queries can be grouped into two classes: class of the one using decomposition-merge-join [12] [13], and the class of those making a direct evaluation [10] [14] [15] [16]. Among algorithms developed in the second class and handling only non-recursive documents, *TreeMatch* algorithm developed by Yao *et al.* [10] has the best time complexity. Generally, in these algorithms, each XML document is manipulated through an index built from an annotation of the latter following the coding (*region coding*) introduced by Li *et al.* [17] where each document node is represented by a (*start*, *end*, *level*) triplet. In the subsequent, we write $Q = (N_q, E_q)$ the exact XML query to be evaluated on the XML document noted $D = (N_d, E_d)$.

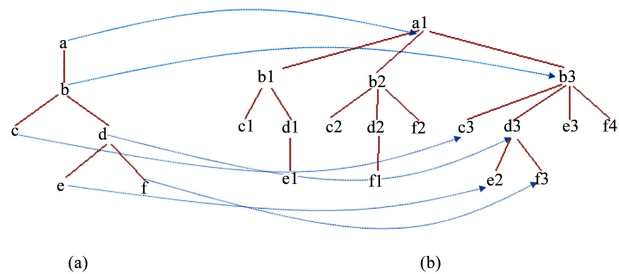

**Figure 1.** Example of the matching (blue lines) of an exact query (a) on an XML document (b).

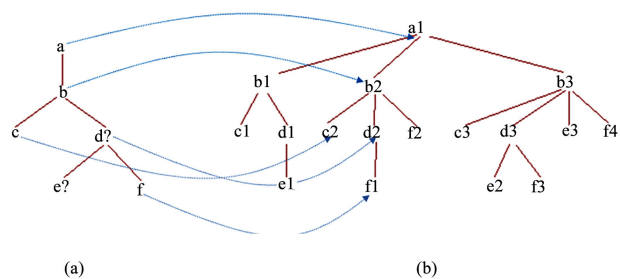

**Figure 2.** Example of matching (blue lines) a preferences query (a) on an XML document (b).

### 2.2.1. Data Structures and Constraints Used in *TreeMatch*

In *TreeMatch*, in order to evaluate query $Q$ on document $D$, a document index based on *region coding* as indicated above is firstly built. From this index, linked lists $T_q$, $q \in N_q$ [6] are retrieved: $T_q$ is a linked list of $q$'s occurrences in $D$; these occurrences are sorted according to the *start* component of the triplets annotating the various occurrences of the node $q$. A stack $S_q$ is associated with each query node $q$; it will store solutions of the evaluation of the *subquery* rooted in $q$.

For each of the $T_q$ lists, a traversing pointer (called *TqCurrent*) is used to locate at any time the $T_q$'s element being processed; it initially points on the first $T_q$'s element. Three primitives are used in *TreeMatch*: *Advance*($Tq$) which makes *TqCurrent* point on the next element of $Tq$, *Isleaf*($q$) which tests if the query node $q$ is a leaf and *NumOfChildren*($q$) which returns the number of children in the query node $q$; we note $q_i \left(i = 0, \cdots, (NumOfChildren(q) - 1)\right)$ a $q$'s child.

An important property of queries handled by *TreeMatch*—which we also use as a hypothesis—is that of No-Self-Containment *i.e.* queries nodes's tags are non-recursive. Indeed, as underlined in [10], for any query $Q = (Nq, Eq)$ satisfying this property, for any internal query's node $q$, any pair of nodes {$x, y$} appearing in $T_q$ have no common descendants. As a result, when executing *TreeMatch*, $T_q$ lists are traversed without going back. Moreover, the matching's result of subquery rooted on $q \in N_q$, can be partition into (disjoint) subsets, each containing an occurrence of $q$. It is this partitioning which allows a compact encoding of the partial solutions (stored in stacks), from which final (global) solution is generated, partition after partition.

### 2.2.2. Principle

Let $Q = (N_q, E_q)$ be an exact query, $D = (N_d, E_d)$ a document. *TreeMatch* [10] algorithm looks for all matchings of $Q$ in $D$ by recursively calling the *find*($q$) function to determine if the current occurrence of $q$ in $Tq$ (given by pointer *TqCurrent*) belongs to a partial solution. Where appropriate, the constituents of the said (partial) solution are compactly encoded in stacks associated with nodes of the subquery rooted in $q$.

Note that, a partial solution is an intermediate result obtained by evaluating a $q$ subquery of $Q$ by matching all the $q$ subquery's nodes (from the occurrence pointed to by *TqCurrent*) with nodes in $D$; it is such that all occurrences of descending nodes of $q$ are descendants of the occurrence of $q$ pointed to by *TqCurrent*: such $D$'s node occurrences are said to be *covered* by the current occurrence of $q$ in $T_q$ [7]. Partial solutions are likely to be completed further to obtain a com-plete solutions; if it cannot be done for a given partial solution, it must be properly removed from various stacks. Thus, when ending the query $Q$ processing, the various (partials) solutions are encoded in the different stacks; all that remains is to browse them appropriately in order to synthesize the final solutions and send them back in a more usable form (a tree for instance): *GenerateSolution*() is the routine that performs this processing in *TreeMatch*.

---

[6]In order to simplify the presentation, as long as there is no ambiguity, in the rest of this paper, we will not make any difference between a node and its label.

[7]In general, when a node *n* belongs to a subtree rooted in *r*, we say that *r covers n*; in this case, considering the *li*'s coding, we have: *r.start* < *n.start* < *r.end*.

## 3. The *TreeMatchPreference* Algorithm

*TreeMatchPreference* is an enriched version of *TreeMatch* that emphasizes on processing induced by the potential presence of preference nodes in a preference query.

### 3.1. Principle

Like *TreeMatch*, *TreeMatchPreference* is recursive. As input it takes a preference query as well as an index (a set of $Tq$ lists) of the document to be queried. It matches the query with the document (through its index) and returns the *best matches*, by recursively calling function *find* ($q$, *TqCurrent*, *maxPosition*)[8], to find all possible matches in the subquery rooted in $q$, from its *TqCurrent* occurrence. The result obtained on each node is encoded in the stack associated with it.

Function *find* ($q$, *TqCurrent*, *maxPosition*) is called to determine if the current occurrence *TqCurrent* of $q$ is a partial solution; its third parameter (*maxPosition*) makes it possible to ensure that in recursive calls, the only occurrences of descending nodes of $q$ that will be taken into account will be exclusively those whose "*end*" component of their region encoding is less than *maxposition*. Indeed, as illustrated in **Figure 3** and **Figure 4**, if we have a query in which $qi$ (child of $q$) is a preference node, we have to distinguish two situations during the matching: the first one is relative to the case in which several occurrences of $qi$ are covered by the same occurrence of $q$ (see **Figure 3**), and the second one concern the case in which two consecutive occurrences of $qi$ are covered by two different occurrences of $q$ (see **Figure 4**). Finally, during the execution of function *find* ($q$, *TqCurrent*, *maxPosition*), the current occurrence *TqCurrent* of $q$ is a partial solution if: either $q$ is a leaf, or, each $q$'s child has a partial solution covered by *TqCurrent*. Moreover, if $q$ is a preference node, *find* ($q$, *TqCurrent*, *maxPosition*) must determine whether there are partial solutions that do not include the current occurrence of $q$: such solutions exist if each of the $q$'s child has at least a partial solution in the subtrees situated between the current $q$'s occurrence (*TqCurrent*) and the one that follows it in $T_q$. In the following, we qualify such partial solutions as *mediocre partial solutions* because they do not contain any occurrences of the (preference) query node being processed.

Generally speaking, during the matching process of a query rooted in $q$, each of its (descending) internal nodes $q_i$ will be process by sequentially applying the same series of operations (grouped in phases, see **Figure 5**) on each of its occurrences. In particular, we distinguish three phases named *phase* 1, *phase* 2 and *phase* 3; only operations in *phase* 2 are applied on occurrences of required nodes. In fact, for a preference node $q_i$, *phase* 1 (resp. *phase* 3) look for the existence of *mediocre partial solutions* not covered by the current occurrence of $q_i$ and belonging to the subtrees preceding it[9] (resp. following it); these occurrences must also be covered by the current occurrence *TqOc* of *q* ($q_i$ is a *q*'s child node). In *phase* 2 however, the processing attempts to find the existence of partial solutions covered by the current occurrence of $q_i$.

---

[8]From the function *find*()'s type, we can already note that it differs from the one used in *TreeMatch*.

By grouping in phases treatments to be performed to match $q_i$ (a *q*'s child node), two main cases (see **Figure 3** and **Figure 4**) can be observed: The first case (see **Figure 3**) concerns the processing of two (many) successive occurrences of $q_i$ (in *Tqi*) covered by the same occurrence of *q*: here, the process performed during the *phase* 3 processing of the first occurrence coincides with that performed during the *phase* 1 processing of the second occurrence. In the second case (see **Figure 4**), two occurrences of $q_i$ are covered by two different occurrences of $q$: here, we must distinguish the *mediocre partial solutions* resulting from the processing in *phase* 3 of the first occurrence of $q_i$ and covered by the first occurrence of $q$, from the *mediocre partial solutions* resulting from the processing in *phase* 1 of the second occurrence of $q_i$ and covered by the second occurrence of $q$. As mentioned before, it is the *maxPosition* parameter of function *find*(), and the fact that lists *Tqi* are traversed without going back that allows to solve this problem.

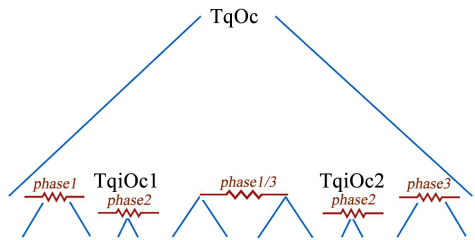

**Figure 3.** Occurrences of $q_i$ covered by the same occurrence of *q*.

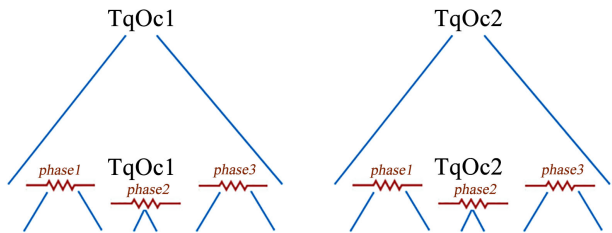

**Figure 4.** Occurrences of *qi* covered by different occurrences of *q*.

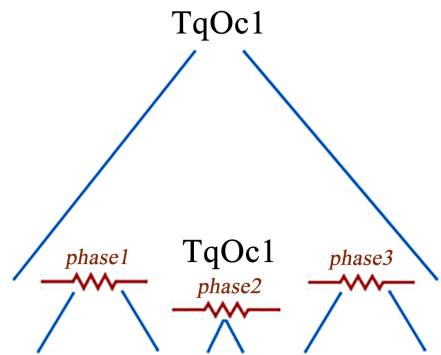

**Figure 5.** An illustration of the three phases to be observed when processing a query node instance.

---

[9] As we will see in function *find*() (**Algorithm 2**), it is the third parameter (*maxPosition*) of *find*() that will discriminate. Indeed, nodes involved in this phase are those whose *end* components of their region encoding are less than *maxPosition*.

## 3.2. Algorithm

In order to facilitate the management of the *mediocre partial solutions* resulting from the introduction of preference nodes, update is made to the structures of the tuples inserted in stacks used in [10].

$q$ being a query node and *prefNode*($q$) (resp. *exactNode*($q$)) designating the set of preferential (resp. required) nodes covered by $q$, the set of $q$'s descending nodes is given by: *descendingNodes*($q$) = *prefNode*($q$) $\cup$ *exactNode*($q$). For instance, for the query sketched in **Figure 2(a)**, we have: *prefNode*($a$) = {$d$, $e$} and *exactNode*($a$) = {$a$, $b$, $c$, $f$}. Let *tupq* be the encoding of a tuple to be pushed in *Sq*. *tupq* consists of five fields and is of the form *tupq* = (*self*, *parent*, *exactnode*, *prefnode*, *pushCount*) in which:

- *self* contains the occurrence tqcurrent of $q$.
- *parent* contains the parent's occurrence of $q$ in $Q$; it covers that contained in *self*.
- *exactNode* contains sequence $d1, \cdots, dk$ of *exactNode*($q$)'s nodes occurrences.
- *prefNode* contains sequence $dk+1, \cdots, dn$ of *prefNode*($q$)'s nodes occurrences.
- *Pushcount* is a counter incremented when pushing in the stack. This field is used as needed by the *cleanStack*() function (**Algorithm 4**) to pop "*Pushcount*" tuples in the current stack when a partial solution cannot be completed to obtain a complete one.

*exactNode* and *prefNode* fields are used exclusively during the generation of the final solution (**Algorithm 3**).

### 3.2.1. The *TreeMatchPreference* Algorithm for Preference Queries

In *TreeMatchPreference* (**Algorithm 1**, line 1), *TqCurrent* initialy point on the first node in $Tq$[10]. Then, if $q$ is a preferred node, the *phase*1 processing is done with the eventual update of the stack associated with q (lines 3-5).

The *phase* 2 is then triggered with the call of function *find*($q$, *TqCurrent*, *TqCurrent.end*) to check if the current $q$'s occurrence (*TqCurrent*) can extend some partial solutions to get complete ones. If so, the stack associated with $q$ is updated accordingly (lines 6 - 7). Then, *TqCurrent* is moved on to the $T_q$'s next item (line 8) to activate the *phase* 3 (lines 9 - 15) and the cycle starts again. Algorithm 1 ends with a call to the function in charge of browsing the different stacks to synthesize the final answer (line 16).

---

[10]In function *find*(), we use the routines *pHead*($Tq$) that return the reference of the first element of $Tq$ and *dropHead*($Tq$) which removes the first element of $Tq$ and return the residual list.

Tuples to be pushed in stacks have the form <*tqcurrent*, *parent_*, *Pushcount*> where, "*tqcurrent*" is the value to be assigned to the *self* field: this value is either the $q$'s current occurrence, or a symbolic value denoted $\Phi$ if $q$ is a preference node; in this case, one must be either in the *phase* 1 or else in the *phase* 3 of the treatment. *parent_* is the value to be assigned to the tuple's *parent* field: this value is the value of the parent node's occurrence of $q$ covering *tqcurrent*, or is equal to the symbolic value "*root*" if this parent must be located beyond the root (Algorithm 1, lines 5 and 12).

### 3.2.2. The Find Algorithm for Preference Queries

The new version of the *find*() function that handles preferences takes three input parameters: *find*(*q*, *TqCurrent*, *maxPosition*) (Algorithm 2); the first parameter $q$ is the current node of the query being evaluated, the second parameter *TqCurrent* is either a $q$'s occurrence, or (in the case where we are in phase 1) the occurrence of his closest ancestor. The third parameter *maxPosition* is an integer used to set an upper limit on the positions of the nodes that can be processed during the call[11].

The different processing phases of $q$'s occurrences are highlighted in Algorithm 2 via comments. Treatments related to *phase* 1 are encoded in lines 7 - 10, while those relating to *phase* 2 are in lines 25 - 27; finally, in lines 29 - 34 we have those relating to *phase* 3. When partial solutions encoded in stacks cannot be extended to obtain a complete solution, they are popped (Algorithm 2, lines 13 - 15) by using function *cleanStack*() (Algorithm 4).

```
Input   : A preference query: a tree;
          An index T (Tq lists) of the document to query;
Output  : An encoding of all matches in the stack associated with the root node
1  TqCurrent =pHead(Tq); Pushcount = 1;
2  while (TqCurrent is not null) do
3      if (q ∈ Npref) then                                      /* Treatment Phase 1 */
4          if (find(q, root, TqCurrent.start) == TRUE) then
5              push(Sq, Φ, root, Pushcount ++);
6      if (find(q, TqCurrent, TqCurrent.end) == TRUE) then      /* Treatment Phase 2 */
7          push(Sq, TqCurrent, root, Pushcount ++);
8      TqCurrent =dropHead(Tq);
9      if (q ∈ Npref ) then                                     /* Treatment Phase 3 */
10         if (TqCurrent is not null) then
11             if (find(q, TqCurrent, TqCurrent.end) == TRUE) then
12                 push(Sq, Φ, root, Pushcount ++);
13         else
14             if (find(q, root, ∞) == TRUE) then
15                 push(Sq, Φ, root, Pushcount ++);
16  GenerateSolution(q);
```

Algorithm 1: The *TreeMatchPreference* algorithm for preference queries.

---

[11]For example, with the call *find*(*q*, *TqCurrent*, 20), only nodes with a value less than 20 in the "*start*" component of their region encoding will be considered.

### 3.3. The Result Generation Process

Function *GenerateSolution*() (**Algorithm 3**) is called to produce an explicit (tree) representation of the final result—consisting of the *best answers* to the query—from the information encoded in the tuples stored in the stacks.

As mentioned in the introduction, intuitively, a match (a tree) will be more preferred if it contains the largest number of preference nodes's occurrences. However, note immediately that two matchings (let's call them *r*1 and *r*2) can be incomparable; this is the case if there are at least two query nodes, *X* and *Y*, such that exactly one of the two matches, *r*1 or *r*2, contains an *X*'s occurrence and in this case, only the other contains a *Y*'s occurrence. It is from this observation that we had the inspiration to use an instrumented version of the *Skyline opera­tor* [11] for the selection of the best answers. In fact, by using this operator, we can define a partial order on the answers so that the best answers are the maxi­mum elements for that order.

Recall that, as defined in RDB, *Skyline operator* [11] allows to select the best n-tuples of a relational table; it can be briefly, presented as follows: let's consider a relational table $R$ with schema $R(P_1, \cdots, P_K, P_{K+1}, \cdots, P_n)$ and two tuples $p = (p_1, \cdots, p_k, p_{k+1}, \cdots, p_n)$ and $q = (q_1, \cdots, q_k, q_{k+1}, \cdots, q_n)$ in *R*. For queries in which preferences relate to fields $P_{k+1}, \cdots, P_n$, we'll say that ***p dominates q*** and we write $p > q$, if the following conditions are met:

1) $p_i = q_i$, for all $i = 1, 2, \cdots, k$ ./* required node occurrences coincide */.

2) $p_i \geq q_i$ for all $i = (k+1), \cdots, n$.

3) there are $i, (k+1) \leq i \leq n$, and $p_i > q_i$.

By considering the so far manipulated data structures, (see Sec. 3.2), we pro­pose a new definition of the *notion of dominance* that is more suitable for com­paring partial solutions produced during the matching; this notion will allow us to easily synthesize the best answers: they are those which are not dominated.

**Definition 1.** *The dimension of a tuple—Let*
$tup_q = (parent, self, d_1, \cdots, d_k, d_{k+1}, \cdots, d_n)$ *be a tuple encoded in a stack. The dimension of* $tup_q$ *noted* $dim(tup_q)$, *is the set of query nodes* $q_{d_{k+1}}, \cdots, q_{d_n}$ *for which* $d_{k+1}, \cdots, d_n$ *are occurrences:* $dim(tup_q) \subseteq prefNode(q)$.

**Definition 2.** *Non dominated Tuple—Let*
$tup_q = (parent, self, d_1, \cdots, d_k, d_{k+1}, \cdots, d_n)$ *and*
$tup'_q = (parent', self', d'_1, \cdots, d'_k, d'_{k+1}, \cdots, d'_m)$ *be two tuples belonging to the same stack. We say that* $tup_q$ *dominates* $tup'_q$ *and we write* $tup_q > tup'_q$, *if* $tup'_q$'s *dimension is included* (*strictly*) *in* $tup_q$'s *dimension:*
$\left( (tup_q > tup'_q) \Leftrightarrow \left( dim(tup'_q) \subset dim(tup_q) \right) \right)$.

Function *GenerateSolution*() (**Algorithm 3**) is similar to its counterpart with the same name in [10]. However, it differs from the latter in that it performs special processing on preference nodes in order to take into account the possible absence of occurrences of such nodes in encoded answers store in stacks. Indeed, among answers stored in stacks associated to preference nodes, only the best (*qualitative preference*) should be retained. The *Skyline operator* provided with

```
Input   : q: a tagged node of the query;
          T_qCurrent: a reference to a q's occurrence in the document;
          maxPosition: an upper bound on the positions of the nodes that can be processed;
Output: A Boolean: True if T_qCurrent is a partial solution, False otherwise.
1  if (isLeaf(q)) return True ;
2  N = NumOfChildren(q);
3  i = 0; PartialSolution = False; PushCount=1;
4  T_qi.Current = pHead(T_qi);
5  while ((PartialSolution == True) OR (T_qi.Current is not null) OR (qi ∈ N_pref)) do
6      if (T_qi.Current is null) OR (T_qi.Current.start > maxPosition) then             /* Treatment Phase 1 */
7          if (qi ∈ N_pref) then
8              if (find(qi, T_qCurrent, maxPosition) == True) then
9                  push(S_qi, Φ, T_qCurrent, PushCount ++);
10                 PartialSolution =True;
11             if (PartialSolution == True) then
12                 i = i + 1;    PartialSolution = False;
13             else
14                 j = 0;
15                 while (j++ < i) do cleanStack(q_j); return False;
16             if (i == N) return True;
17      else
18          if (T_qCurrent.start < T_qCurrent.start) then
19              T_qCurrent = dropHead(T_qi);
20          else                                                                         /* Treatment Phase 1 + 2 + 3 */
21              if (qi ∈ N_pref) then                                                     /* Treatment Phase 1 */
22                  if (find(qi, T_qCurrent, T_qCurrent.start) == True) then              /* see Phase 1 on figure 5 */
23                      push(S_qi, Φ, T_qCurrent, PushCount ++);
24                      PartialSolution = True;
25              if (find(qi, T_qCurrent, T_qCurrent.end) == True) then                    /* see Phase 2 on figure 5 */
26                  push(S_qi, T_qCurrent, T_qCurrent, PushCount ++);
27                  PartialSolution = True;
28              T_qCurrent = dropHead(T_qi);
29              if (qi ∈ N_pref) then                                                     /* Treatment Phase 3 */
30                  if (T_qCurrent is not null) then
31                      newMaxPosition = min(maxPosition, T_qCurrent.start);
32                      if (find(qi, T_qCurrent, noulleMaxPosition) == True) then
33                          push(S_qi, Φ, T_qCurrent, PushCount ++);
34                          PartialSolution = True;

35 return False;
```

**Algorithm 2:** The find algorithm for preference queries.

the dominance relation defined above makes it possible to make this selection: *an answer (a tuple) is retained only if it is not dominated by any other*. Function *filterSkylineSolution*() called in **Algorithm 3** is used to incrementally make this selection.

More precisely, function *GenerateSolution*() (**Algorithm 3**) performs a preorder traversal of different stacks associated with nodes of the query tree. At each node, it realizes the equijoin of the associated stack and those of her children and stores the non dominated tuples in the same stack (overwrite). Finally, query's answer is synthesized from the leaves of the query tree and is stored in the stack associated with the root node of the query tree.

```
Input  : The query tree decorated with stacks;
Output: An explicit representation of the final result of the query.
1  N = NumOfChildren(q);                           /* q is the root node of the query tree */
2  i = 0;
3  P = creerPile();
4  while i++ < N do
5      generateSolution(q_i);
6      S_q = JoinAndFilter(S_q, S_qi) ;
7  P = filterSkylineSolution(S_q);                  /* Selecting non-dominated answers in S_q */
8  return P;
```

**Algorithm 3:** The *GenerateSolution* algorithm for preference queries.

### 3.4. Illustration

This subsection is devoted to a completely illustrated presentation (from the data initialization to the generation of the final solution) of the matching of a preference query (see Figure 6) on an XML data source (see Figure 7). An index $T$ (the Tq lists, Figure 8) is build from the XML data source, and the evaluation function *TreeMatchPreference* ($a$, $T$) is called to match the query with the data source. Notice that, parameter "$a$" used in the call is the root node of the query (see Figure 6).

*TreeMatchPreference* starts by examining whether the first occurrence "$a1$" of "$a$" in $T_a$ is a solution to the query rooted in "$a$". More precisely, the goal is to determine (see Figure 9) if "$a1$" is the root of a subtree consisting of well-structured occurrences of $b$, $c$, $f$ and possibly $d$ and $e$? If so, the tuple < $a1$, $root$, 1> must be pushed in $S_a$ (the stack associated with the query node $a$).

---

**Input** : q: a query node;
1 N = NumOfChildren(q);
2 i = 0;
3 **while** *(i++ < N)* **do**
4     cleanStack($q_i$);
5 compt = $S_q \rightarrow$ top.PushCount;
6 **while** *((S_q is not empty) AND (compt > 0))* **do**
7     *Pop($S_q$)* ;
8     compt- -;

---

**Algorithm 4:** The *cleanStack* algorithm.

---

**Input** : Two queries nodes $q$ and $q_i$, $q_i$ is the $i^{th}$ son of $q$ ;
**Output**: The stack containing the equijoin of stacks $S_q$ and $S_{q_i}$ respectively associated with $q$ and $q_i$.
1 **while** *isEmptyStack($S_q$) == False* **do**
2     P = newStack(); P' = newStack(); tp = pop($S_q$);
3     **while** *isEmptyStack($S_{q_i}$) == False* **do**
4        tpi = pop($S_{q_i}$);
5        **if** *((tp.self === Φ)AND(tp.parent == tpi.parent))* **then**
6           PartialSolution parSol = (tp.self, tp.parent, concat[$prefNode(tp)$, $prefNode(tp_i)$], concat[$prefNode(tp)$, $prefNode(tp_i)$]);
7           push(P, parSol);
8        **else**
9           **if** *(tp.self == tp.parent)* **then**
10              PartialSolution parSol = (tp.self, tp.parent, concat[$prefNode(tp)$, $prefNode(tp_i)$], concat[$prefNode(tp)$, $prefNode(tp_i)$]);
11              push(P, parSol);
12           **else**
13              push(P', $tp_i$);
14     $S_{q_i} = P'$ ;
15 **return** *filterSkylineSolution(P)*;

---

**Algorithm 5:** The *joinAndFilter* algorithm.

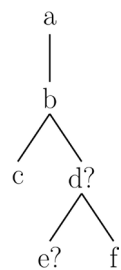

**Figure 6.** A tree representation of a preference query used to query the data source of Figure 7.

---

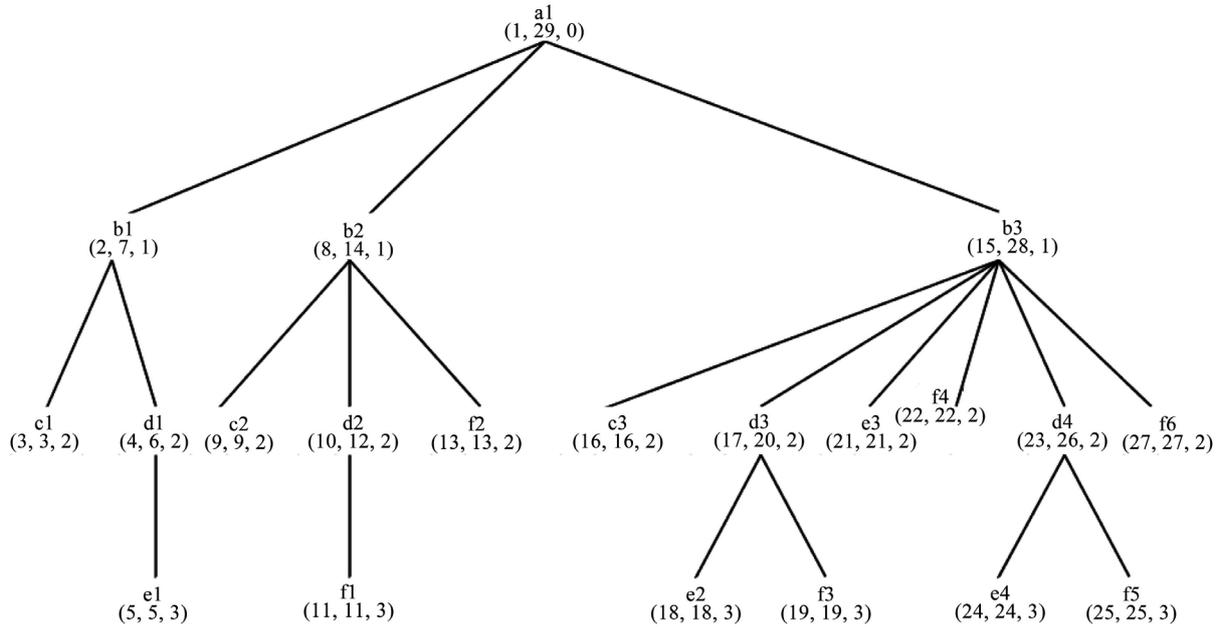

**Figure 7.** Region coding labelling of the nodes of an update of **Figure 1(b)**.

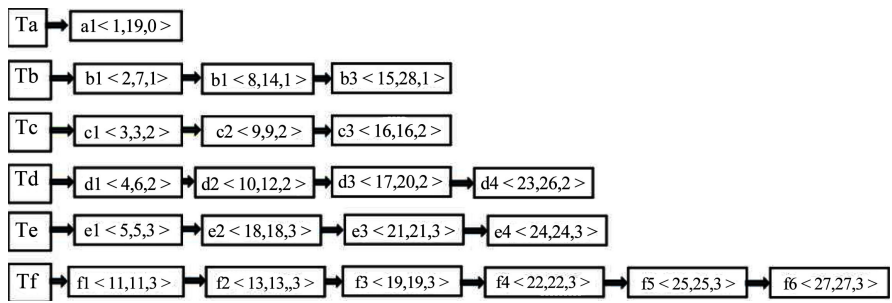

**Figure 8.** $T_q$ obtained from the data source of **Figure 7**.

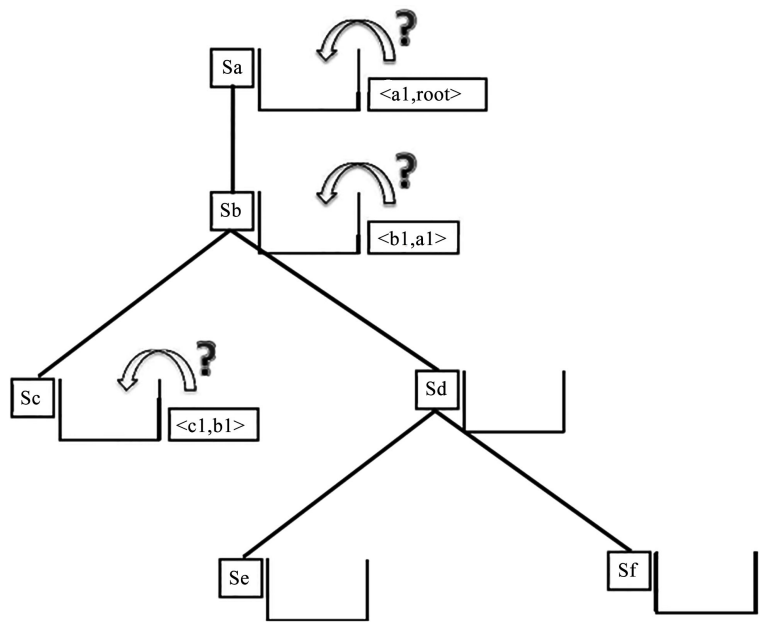

**Figure 9.** Evaluation of the query node *a*.

Knowing that node "*a*" has a son (node "*b*") and that *TreeMatchPreference* make a call to the recursive function (*find*()) when matching, the previous question is renewed on the current occurrence "*b*1" of "*b*" to examine if "*b*1" is a partial solution[12]? This question is equivalent to this one: "*should we push tuple $< b1, a1>$ in $S_b$?*". Like before, due to the function *find*() recursion, the question "*should we push tuple $< c1, b1>$ in $S_c$?*" is examined (see **Figure 9**) and the answer is "Yes" since, *c* is a leaf node and *c*1 is covered by *b*1; tuple $< c1, b1>$ is then pushed in $S_c$ (see **Figure 10(a)**).

After tuple $< c1, b1>$'s stacking, $T_c$'s traversing pointer is moved to the next occurrence of *c*, "*c*2" then becomes the new current occurrence of "*c*". Since *c*2 is not covered by "*b*1", evaluation continues on "*d*" (the second son of "*b*"); the current occurrence of "*d*" in $T_d$ is "*d*1". Since *d* is a preference node, *TreeMatchPreference* examine if tuple $<\Phi, b1>$ can be pushed in $S_d$? This corresponds to the execution of the so-called *phase* 1 (**Algorithm 2**, line 8). As above, the recursion of *find*() and the fact that *e* is a preferential node must lead to another test: "*should we push tuple $<\Phi, b1>$ in $S_e$?*" (see **Figure 10(b)**). Note that in the tuple $<\Phi, b1>$, the attribute *parent* is set to *b*1 (and *self* = $\Phi$); this reflects the fact that the parent of *e* does not have any occurrences in the partial solution under construction[13].

The test "*push $<\Phi, b1>$ in $S_e$?*" is positive since *e* is a leaf (**Algorithm 2**, line 1); $<\Phi, b1, 1>$ tuple is then stacked (see **Figure 11(a)**) and the $T_e$'s traversing pointer is moved to "*e*2". Since "*e*2" is not covered by "*d*1", the processing continues with the evaluation of *f* (the next child of "*d*"), *f*1 being its current occurrence. The test "*push $< f1, b1>$ in $S_f$?*" is performed with a negative answer since *f*1 is not covered by *b*1: $<f1, b1>$ can't be pushed in $S_f$ (see **Figure 11(a)**). The negative answer to the test "*push $< f1, b1>$ in $S_f$?*" imposes the deletion (by using function *cleanStack*) of one tuple (*pushCount* = 1) in $S_e$ which is "linked" to "*b*1" in $S_d$ (see **Figure 11(b)**).

After going through all the lists $T_q$ ( $q \in \{a, b, c, d, e, f\}$ ), *TreeMatchPreference* will have stored in each stack (see **Figure 12**) a compact encoding of all solutions (tuples) of the query.

Function *generateSolution*() is then called to produce the non dominated tuples. To do this, it merges and cleans the contents of stacks by performing at each node the equijoin of the stack associated with it and that of its children; only non dominated tuples are retained. Function *joinAndFilter*() (**Algorithm 5**) is used for this purpose.

**Figure 13** shows the executing result of function *generateSolution*() when it is run over the stacks sketched in **Figure 12**. *generateSolution*() being recursive and performing a pre-order traversal of the query tree, it is invoked by providing it with the root node of the query. The first merge is made between the $S_b$ and $S_c$ stacks to provide the new $S_b$ stack (dotted in **Figure 13**). The second merge is made between the $S_c$ and $S_d$ stacks to obtain the stack sketched in **Figure 14(a)**. Note that each tuple of this stack is followed by a pair of binaries $(x, y)$ in which, $x$ (resp. $y$) allows us to know if the tuple contains ("1") or not ("0") an occurrence of a preference node $e$ (resp. $d$). For instance, for a stack entry containing $<d4, b3, e4> (1, 1)$, the pair $(x, y)$ is equal to $(1, 1)$; this means that tuple $< d4, b3, e4>$ contain an $e$'s occurrence ($e4$ for this case) and a $d$'s occurrence ($d4$ for this case). On the other hand, $<\Phi, b3, e3> (1, 0)$ indicates the presence ($x =$ "1") of an occurrence of $e$ ($e3$ in this case) and no occurrence ($y =$ "0") of $d$. A tuple that does not have any occurrences of the preference nodes will have this pair equal to $(0, 0)$.

---

[12]Remember that a $b_i$ occurrence of *b* is a partial solution if and only if there exists an occurrence of each of its sons (here *c* and *d*) which are sons of $b_i$ and which are themselves partial solutions.

[13]We are evaluating a *mediocre partial solution* (see Sec. 3.1) which does not include a *d* occurrence. We will then examine if there exist a solution which does not include neither a *d* occurrence nor an *e* occurrence by considering the tuple $<\Phi, d1>$ in which $\Phi$ materializes the absence of an occurrence of *e* in the investigated solution. This confirms the fact that *find*() evaluates a preference node by considering two alternatives: *absent* or *present*, respectively corresponding to treatments grouped in *phase* 1 and *phase* 2 in **Algorithm 2**.

After merging $S_d$ and $S_e$, the filter function is applied to the result to remove duplicate tuples as well as to extract those that are not dominated. **Figure 14(b)** shows the resulting stack; it gives an illustration (on the portion noted [4]) of the extraction of the not dominated tuple $< d4, b3, e4>$ among tuples $\{<d4, b3>, <d4, b3, e4>\}$. The stack of the **Figure 14(b)** is finally the one associated with node "$d$"; it is subsequently merged with $S_f$. Proceeding as for $S_e$, we have the new stack $S_{df}$ (see **Figure 15(a)**), and finally that of $S_{af}$ (see **Figure 16(b)**) which contains only not dominated tuples: these are the best answers sought.

## 4. Implementation and Experimental Study of *TreeMatchPreference*

In the previous section, we described *TreeMatchPreference* together with functions on which it relies upon when evaluating a query. In order to facilitate its

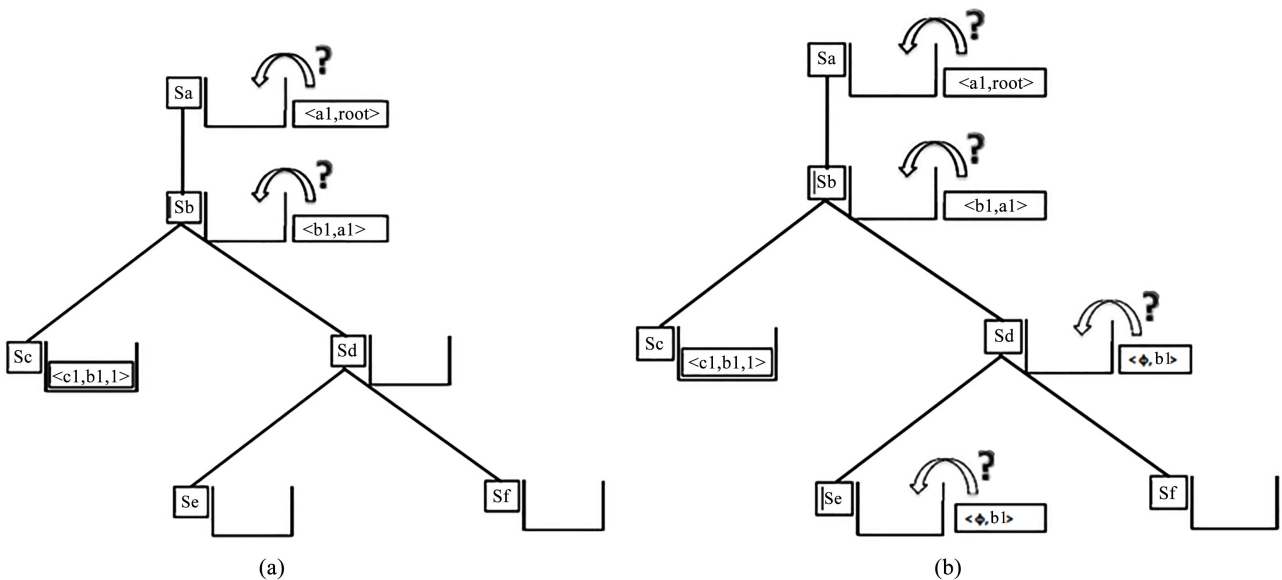

(a)                                     (b)

**Figure 10.** Pushing $<c1, b1>$ in $S_c$ (a) and Evaluation of the preference query node $d$ (b).

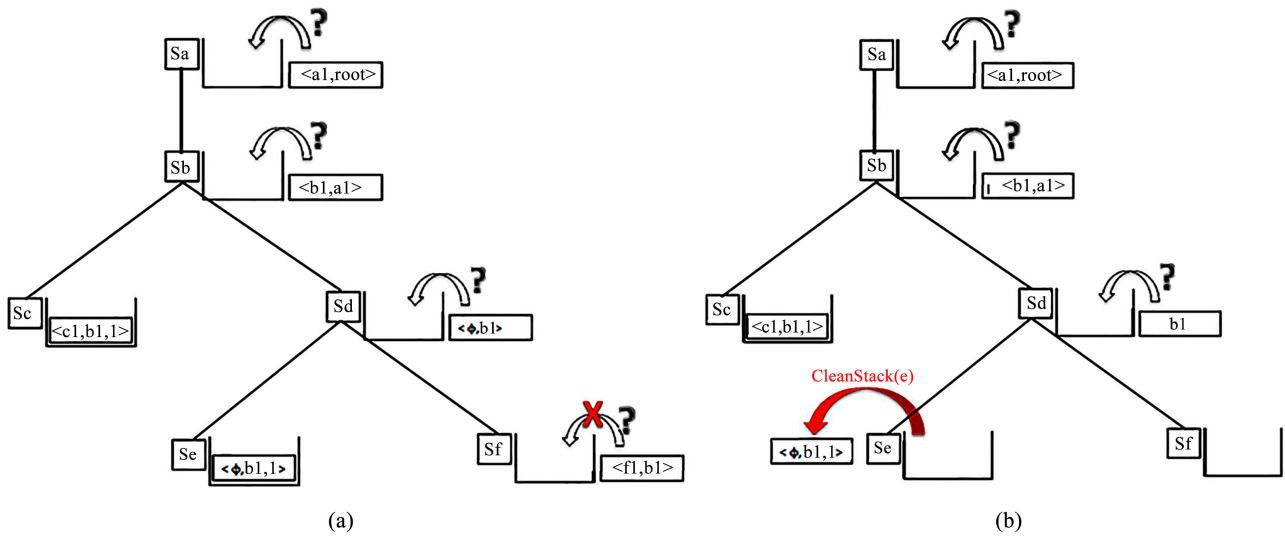

**Figure 11.** Pushing $<\Phi, b1>$ in $S_c$ and negative answer for test of pushing ($<f1, b1> \ in \ S_f$) (a) and pop of $<\Phi, b1>$ from $S_c$ (b).

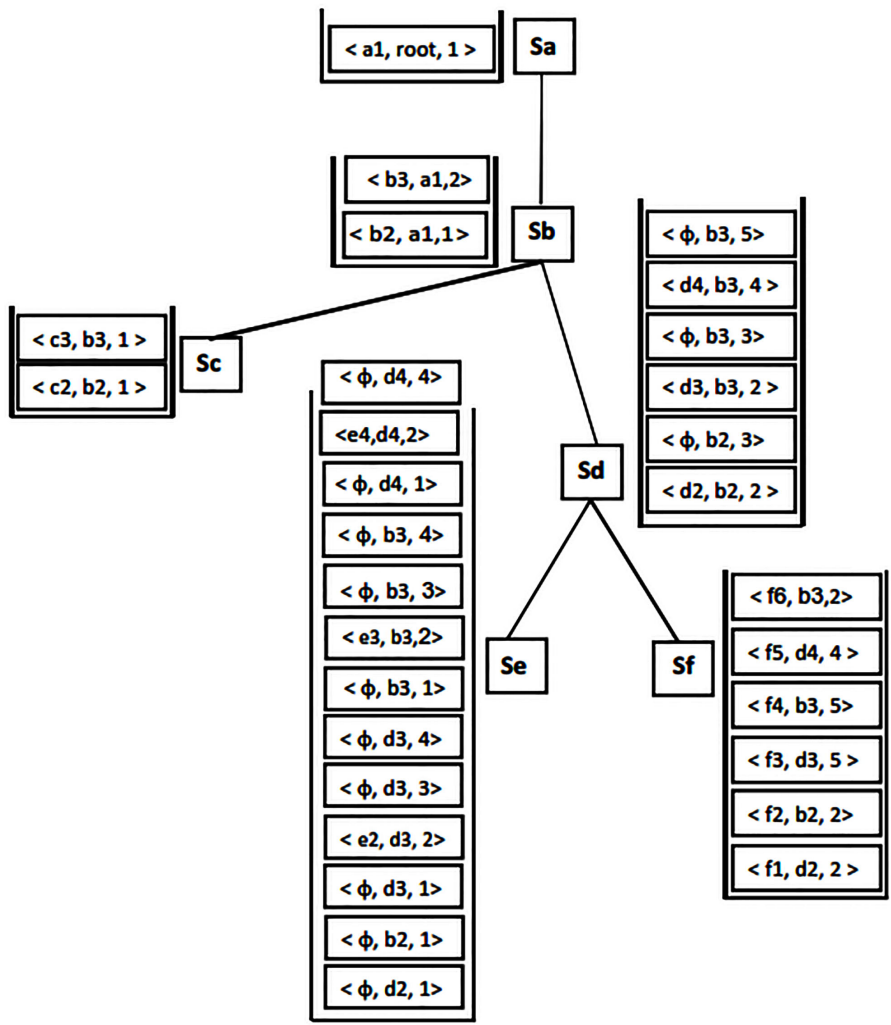

**Figure 12.** Stacks storing the encoding of all matches of the query of **Figure 6** with the data source of **Figure 7**.

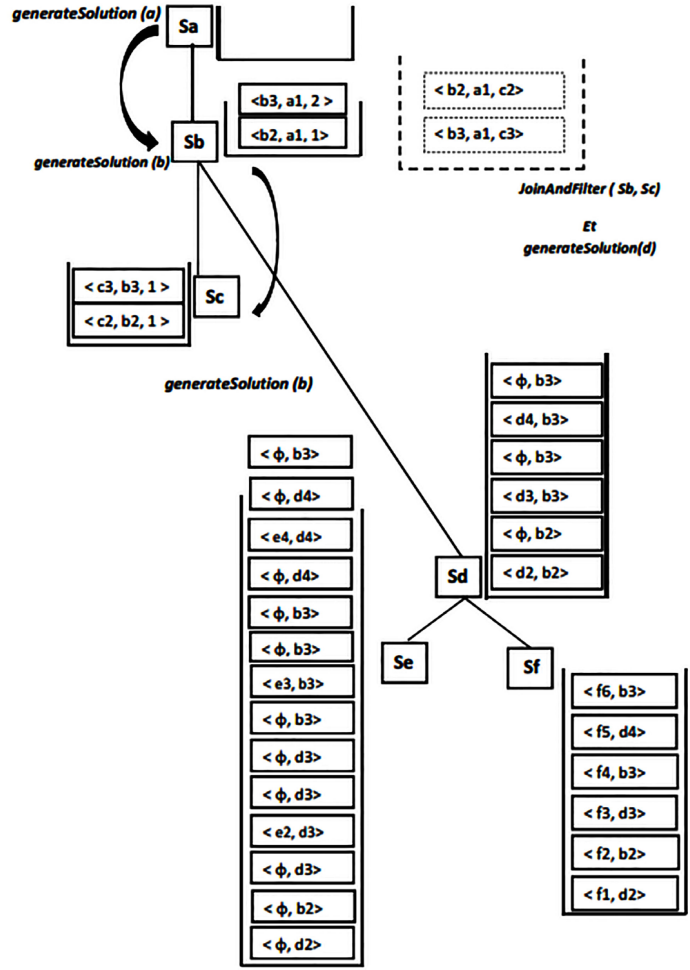

**Figure 13.** Merging $S_d$ and $S_e$ followed by $S_d$'s filtering.

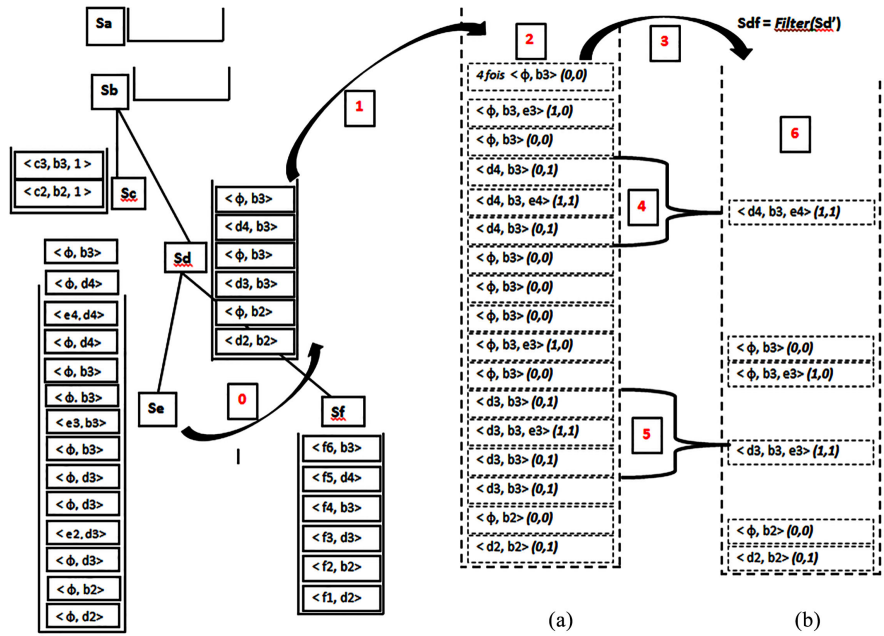

**Figure 14.** Merging $S_d$ and $S_e$ followed by filtering.

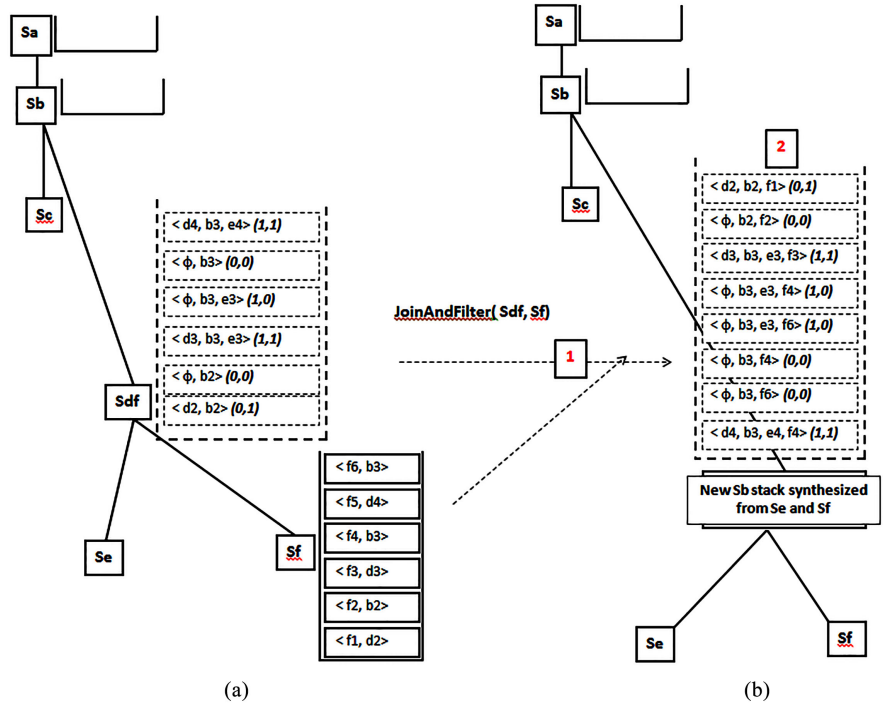

(a)

(b)

**Figure 15.** Merging $S_b$ and $S_c$ followed by filtering.

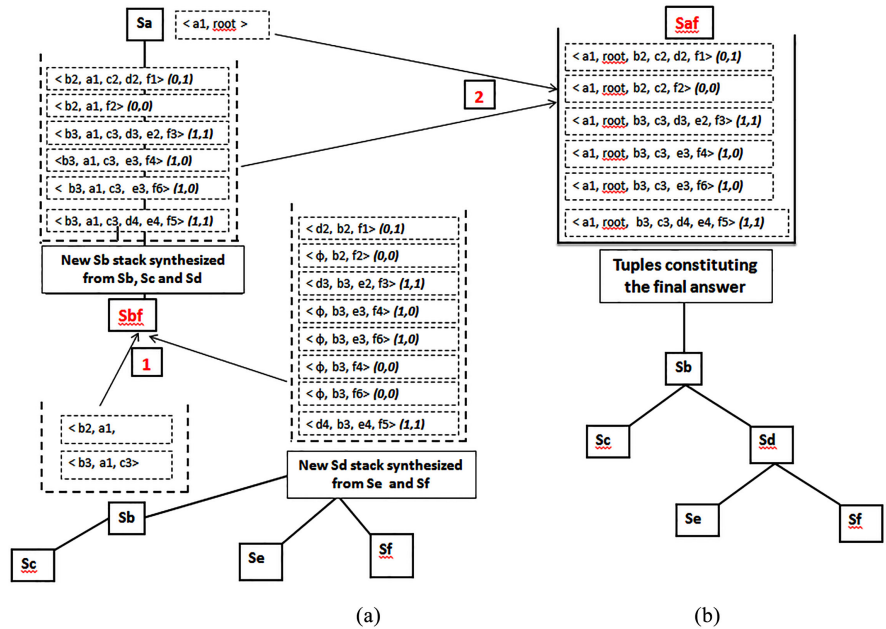

(a)

(b)

**Figure 16.** Merging $S_{df}$ and $S_b$ (a), merging $S_{bf}$ and $S_a$ (b) followed by filtering.

experimentation with well-known and recommended XML DBs for testing XML query evaluation algorithms, we have developed a platform for the expression and execution of preference queries; it allows us to select an XML data source, define or select an existing query, apply a query to a data source, and then view the best answers resulting from the match. Finally, this platform allowed us to easily experiment with the proposed algorithm by using it for the evaluation of many preference queries on various XML data sources. A summary of the work done for this purpose is presented in the following two subsections.

## 4.1. XML Databases and Sample Queries

- *SigmodRecord.xml* contains an index set of SIGMOD registry items (*http://www.dia.uniroma3.it/Araneus/Sigmod*). The following queries have been applied to this file:

1) "*SigmodRecord[/issue[/volume?/15?]/articles/article/authors/author/Sophie Cluet]*": search all articles of author "Sophie Cluet" belonging to the fifteenth volume's publication; the fact of belonging to the fifteenth volume of publications is preferential.

2) "*/authors[/author/Sergey Brin]/author/Rajeev Meitwani?*": search all articles written by authors "Sergey Brin" and "Rajeev Meitwani". In the best case, the returned results will have these two authors and in the worst case, only those of "Sergey Brin" will be returned, because "Rajeev Meitwani" is preferential.

- *part.xml* contains information about products of the Transaction Processing Performance Council (TPC-*http://www.tpc.org/tpch/*). The following queries have been applied to this file:

1) "*table[/T[/INT/P_BRAND/Brand#13?]/P_CONTAINER/SM_CASE]*": this query looks for products from the "SM_CASE" container with a preference over those belonging to the "Brand#13" brand.

2) "*table[/T[/P_TYPE/LARGE?][/P_SIZE/10]/P_CONTAINER/MED_BOX?]*": this query searches for size 10 products in any container with a preference for those from the "MED_BOX" container.

- The last experiment file *mydata.xml* is not part of a repository. It was created by following the XML document fragment of **Figure 7** and two queries ("*a[/b[/c]/d?[/e?]/f]*" and "*a[/b[/c]/d?[/e]/f]*") were executed on it.

## 4.2. An Overview of the Tool Developed for Executing Preference Queries

The main window of the developed prototype is divided into four panels (see **Figure 17**):

- The upper left panel is used to select the data source to query.
- The lower left panel is used to select an existing query or to edit a new one.
- The top right panel is used to display the best answers.
- The lower right panel displays the tree structure of a selected answer in the upper right panel.

For instance, to query *mydata.xml* data source, click on the "*select file*" button (see **Figure 17**) to select the file to query. Once the file is selected, it is parsed and an index based on region coding is generated. Then, a corresponding tree representation is displayed in the dedicated panel, followed by loading existing queries (related to the selected file) into a drop-down menu (from the same panel). Since these loaded queries are editable, it is possible for a user to completely (re)formulate a new one.

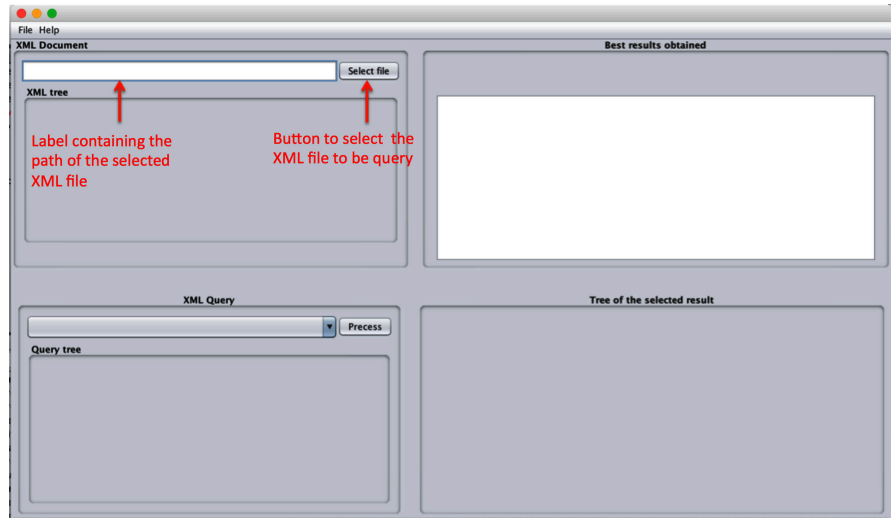

**Figure 17.** The GUI of the tool developed for experimenting the evaluation of preference queries.

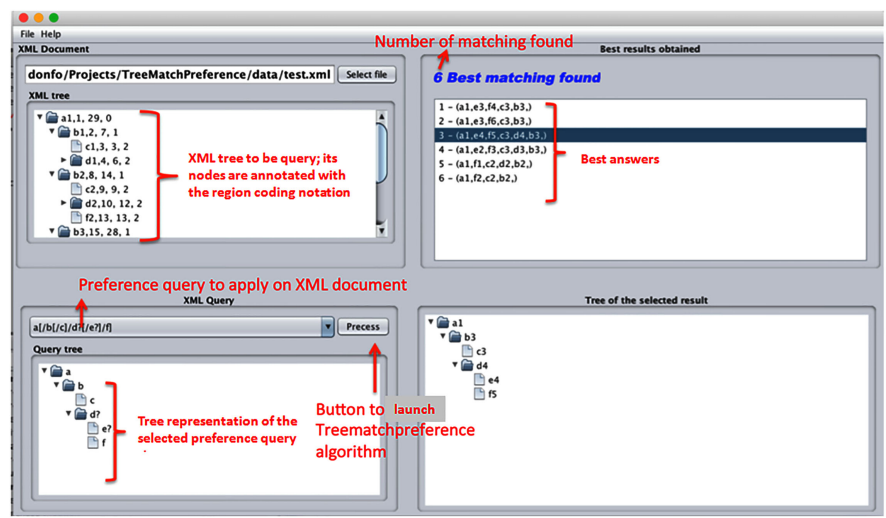

**Figure 18.** A screenshot sketching the matching process.

When selecting an element (a query) in this drop-down menu, its tree representation is drawn just below. Right next to the drop-down menu, the "*Process*" button is used to launch the document querying process. Best matches found are displayed on the top left panel (see **Figure 18**) and the selection of one of them causes the display on the down left panel of its tree representation (see **Figure 18**).

## 5. Conclusions

This article focused on importing the concept of preference queries into the XML community. For this purpose, we have proposed *TreeMatchPreference* which is an algorithm (inspired by *TreeMatch* [10]), to evaluate preference queries on an XML data source. The query language used is a subset of that proposed by Cohen and Shiloach [9] in which we have retained only structural pre-ferences.

Inspired by Skyline techniques used in RBD for evaluating preference queries [18] [19], we proposed a so-called *SkyTrees*. *SkyTrees* is the set of best (non-dominated) answers obtained when using *TreeMatchPreference* to match a preference query on an XML data source.

A short-term work will consist of the full analytical study of the proposed algorithm and its extension to the processing of value-based preference queries.

We also plan to explore a presentation of the proposed algorithms following a grammatical approach inspired by that adopted by Bouchou *et al.* [20] for XML integrity constraint validation: the expected gain is the fluidity of the presentation (algorithm design would be more modular) and the reduction of the query's tree traversing time.

## Acknowledgements

Authors warmly thank ***Beatrice Bouchou*** for suggesting the theme discussed in this paper.

## Conflicts of Interest

The authors declare no conflicts of interest regarding the publication of this paper.